\documentclass{epl}

\title{Intermittent generalized synchronization in unidirectionally
coupled chaotic oscillators} \shorttitle{Intermitted generalized
synchronization\dots}
\author{A.E.~Hramov \and A.A.~Koronovskii}
\institute{Faculty of Nonlinear Processes, Saratov State
University, Astrakhanskaya, 83, Saratov, 410012, Russia}



\usepackage{amssymb}
\begin{document}

\maketitle

\begin{abstract}
A new behavior type of unidirectionally coupled chaotic
oscillators near the generalized synchronization transition has
been detected. It has been shown that the generalized
synchronization appearance is preceded by the intermitted
behavior: close to threshold parameter value the coupled chaotic
systems demonstrate the generalized synchronization most of the
time, but there are time intervals during which the synchronized
oscillations are interrupted by non-synchronous bursts. This type
of the system behavior has been called intermitted generalized
synchronization (IGS) by analogy with intermitted lag
synchronization (ILS) [Phys. Rev. E \textbf{62}, 7497 (2000)].
\end{abstract}

Synchronization~\cite{Pikovsky:2002_SynhroBook,
Anshchenko:2001_SynhroBook} of interacting chaotic oscillators is
one of the fundamental phenomena of nonlinear dynamics. Recently
several types of chaotic synchronization have been observed in
coupled nonlinear oscillators. These are phase synchronization
(PS)
\cite{Pikovsky:2000_SynchroReview,%
Rosenblum:1996_PhaseSynchro,Osipov:2003_3TypesTransitions}, lag
synchronization (LS) \cite{Rosenblum:1997_LagSynchro,Zhigang:2000_GSversusPS,%
Taherion:1999_LagSynchro}, complete synchronization (CS) \cite{Pecora:1990_ChaosSynchro,%
Pecora:1991_ChaosSynchro,Murali:1994_SynchroIdenticalSyst,%
Murali:1993_SignalTransmission} and generalized synchronization
(GS) \cite{Rulkov:1995_GeneralSynchro,Kocarev:1996_GS,%
Pyragas:1996_WeakAndStrongSynchro}. It is important to note that
GS may also takes place in the non-oscillatory chaotic systems
(see \cite{Boccaletti:2002_SynchroPhysReport} for detail). All
synchronization types are interrelated (see for
detail~\cite{Parlitz:1996_PhaseSynchroExperimental,%
Rulkov:1995_GeneralSynchro,Zhigang:2000_GSversusPS}), but the
relationship between them is not completely clarified yet. In
recent
works~\cite{Brown:2000_ChaosSynchro,Boccaletti:2001_UnifingSynchro,%
Hramov:2004_Chaos,alkor:2004_JETPLetters_TSS} it has been shown
that all these synchronization types may be considered from the
common point of view as different manifestations of one universal
phenomenon.

It has also been found that onsets of phase and lag
synchronization types are preceded by intermittent behavior. Close
to the threshold parameter value the coupled chaotic systems
demonstrate synchronized dynamics most of the time, but there are
time intervals during which the synchronized oscillations are
interrupted by non-synchronous behavior. These pre--transitional
intermittencies have been described in
\cite{Rosenblum:1997_LagSynchro,Boccaletti:2000_IntermitLagSynchro,Zhan:2002_ILS}
for the case of lag
synchronization and in \cite{Pikovsky:1997_EyeletIntermitt,%
Pikovsky:1997_PhaseSynchro_UPOs,Lee:1998:PhaseJumps} for phase
synchronization, respectively.

Due to the existence of unifying framework of coupled chaotic
oscillators synchronization one can also expect the intermitted
behavior at the threshold of the generalized synchronization
appearance. In this work we consider the behavior of coupled
chaotic oscillators close to the coupling parameter value
corresponding to the onset of generalized synchronization regime.
As it will be shown below the generalized synchronization
appearance is also preceded by the intermitted behavior in the
same way as the phase synchronization and lag synchronization
regimes do. This type of system behavior we called
\textit{intermitted generalized synchronization} (IGS) by analogy
with intermitted lag synchronization
(see~\cite{Boccaletti:2000_IntermitLagSynchro}).

The \textit{generalized synchronization} (GS)
\cite{Rulkov:1995_GeneralSynchro,Kocarev:1996_GS,%
Pyragas:1996_WeakAndStrongSynchro} regime introduced for drive
$\mathbf{x}(t)$ and response $\mathbf{u}(t)$ systems means that
there is some functional relation between unidirectionally coupled
chaotic oscillators, i.e.
$\mathbf{u}(t)=\mathbf{F}[\mathbf{x}(t)]$. This functional
relation $\mathbf{F}[\cdot]$ can be very complicated, but there
are several methods to detect the synchronized behavior of coupled
chaotic oscillators, e.g., the method of nearest
neighbors~\cite{Rulkov:1995_GeneralSynchro,Pecora:1995_statistics}.
In our work we have used the auxiliary system approach proposed
in~\cite{Rulkov:1996_AuxiliarySystem}. In this case we consider
the dynamics of drive
\begin{equation}
\dot\mathbf{x}(t)=\mathbf{H}(\mathbf{x}(t)) \label{eq:drive}
\end{equation}
and response
\begin{equation}
\dot\mathbf{u}(t)=\mathbf{G}(\mathbf{u}(t),\mathbf{g},\mathbf{x}(t))
\label{eq:response}
\end{equation}
systems. The vector $\mathbf{g}$ characterizes the coupling of the
drive $\mathbf{x}(t)$ and response $\mathbf{u}(t)$ systems. At the
same time we also consider the dynamics of the auxiliary system
\begin{equation}
\dot\mathbf{v}(t)=\mathbf{G}(\mathbf{v}(t),\mathbf{g},\mathbf{x}(t))
\label{eq:auxiliary}
\end{equation}
which is identical to the response system $\mathbf{u}(t)$ but
starts with the other initial conditions, i.e.,
$\mathbf{u}(t_0)\neq \mathbf{v}(t_0)$. In the absence of
generalized synchronization between the drive $\mathbf{x}(t)$ and
response $\mathbf{u}(t)$ systems, the phase trajectories of the
response $\mathbf{u}(t)$ and auxiliary $\mathbf{v}(t)$ systems
share the same chaotic attractor but are otherwise unrelated. In
the case of generalized synchronization the orbits of response
$\mathbf{u}(t)$ and auxiliary $\mathbf{v}(t)$ systems become
identical after the transient dies out due to the generalized
synchronization relations
$\mathbf{u}(t)=\mathbf{F}[\mathbf{x}(t)]$ and
$\mathbf{v}(t)=\mathbf{F}[\mathbf{x}(t)]$. It is obvious that in
the case of generalized synchronization the condition
$\mathbf{u}(t)=\mathbf{v}(t)$ should be satisfied and the identity
of response and auxiliary systems is a much simpler criterion to
test the presence of generalized synchronization rather than
detection of unknown functional relationship $\mathbf{F}[\cdot]$.

\begin{figure}[tb]
\centerline{\includegraphics*[scale=0.35]{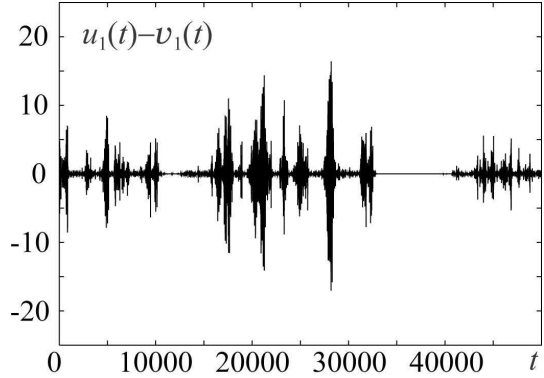}} \caption{The
dependence of difference between the coordinates of response and
auxiliary systems $u_1(t)-v_1(t)$ on time. The coupling parameter
between drive and response systems has been selected as
$\varepsilon=0.106$ \label{fgr:RosslersIGS}}
\end{figure}

For the coupling parameter values close to the onset of
generalized synchronization the drive and response systems
demonstrate the regime of intermitted generalized synchronization.
In this  case the generalized synchronization regime between drive
and response systems has been observed most of the time and at
that time the vector states of response $\mathbf{u}(t)$ and
auxiliary $\mathbf{v}(t)$ systems coincide with each other.
Nevertheless, there are time intervals when the drive and response
systems behave unrelatedly. Therefore, during these time intervals
the vector states of response $\mathbf{u}(t)$ and auxiliary
$\mathbf{v}(t)$ systems differ from each other. If we plot the
difference between vector states of response and auxiliary systems
$|\mathbf{u}(t)-\mathbf{v}(t)|$ versus time variable $t$ than the
graph will be obtained where the laminar phases corresponding to
the generalized synchronous behavior alternate with the turbulent
ones when the generalized synchronization is not detected. In
other words, in the case of IGS for some time intervals one can
detect the presence of functional relation $\mathbf{F}[\cdot]$
between the drive and response systems whereas for the other time
intervals such functional relation does not exist.

Let us consider the dynamics of two unidirectionally coupled
R\"ossler systems
\begin{equation}
\begin{array}{lll}
\dot{x}_1=-\omega_1x_2-x_3& &\dot{u}_1=-\omega_2u_2-u_3+\varepsilon(x_1-u_1)\\
\dot{x}_2=\omega_1x_1+ax_2&\qquad\mbox{and}\qquad &\dot{u}_2=\omega_2u_1+au_2\\
\dot{x}_3=p+x_3(x_1-c)& &\dot{u}_3=p+u_3(u_1-c),
\end{array}
\label{eq:Rslr_drive}
\end{equation}
where $\mathbf{x}=(x_1,x_2,x_3)^T$ and
$\mathbf{u}=(u_1,u_2,u_3)^T$ are the vector states of drive and
response systems, respectively. The control parameter values have
been chosen as $\omega_1=0.99$, $\omega_2=0.95$, $a=0.15$,
$p=0.2$, $c=10$, the parameter $\varepsilon$ characterizes the
coupling strength between the considered systems. At the same time
the behavior of auxiliary system $\mathbf{v}(t)$ described also by
equation (\ref{eq:Rslr_drive}) should be considered in order to
detect the presence of the generalized synchronization regime. As
it follows from our investigation PS arises at
$\varepsilon_{PS}\approx0.116$, GS takes place at $\varepsilon$ at
$\varepsilon_{c}\approx 0.11$, and LS can be observed for
$\varepsilon\geq\varepsilon_{LS}\approx0.36$. So, GS for the
coupling strength exceeding slightly the critical value
$\varepsilon_{c}$ is the week
synchronization~\cite{Boccaletti:2002_SynchroPhysReport,%
Pyragas:1996_WeakAndStrongSynchro} and, therefore, the
relationship $\mathbf{F}[\cdot]$ demonstrates fractal properties.
For $\varepsilon\geq\varepsilon_{LS}$ GS becomes the strong
synchronization as well as $\mathbf{F}[\cdot]$ begins being
smooth. In this case the behavior of the drive and response
systems becomes identical but shifted in time versus each other.
Close to the onset of the GS the regime of IGS has been detected.
The difference between the coordinates of response and auxiliary
R\"ossler systems has been shown in Fig.~\ref{fgr:RosslersIGS}.

\begin{figure}[tb]
\centerline{\includegraphics*[scale=0.35]{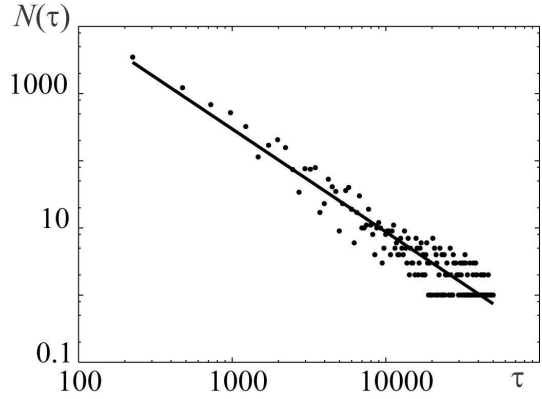}} \caption{The
statistical distribution of laminar phases and its approximation
$N(\tau)\sim \tau^{-3/2}$ in log-log scale. The coupling parameter
between drive and response systems has been selected as
${\varepsilon=0.106}$, the value of the threshold ${\Delta=0.1}$
\label{fgr:DistribIGS}}
\end{figure}

The time intervals where the difference $u_1(t)-v_1(t)$ is close
to zero correspond to the laminar phases of coupled oscillator
dynamics when the presence of functional relation
$\mathbf{F}[\cdot]$ between drive and response systems can be
detected. Alternatively, the irregular bursts indicate breaking of
the relationship $\mathbf{F}[\cdot]$. It is interesting to note
that amplitude of irregular bursts is large enough and comparable
with the size of chaotic attractor of R\"ossler system.

\begin{figure}[tb]
\centerline{\includegraphics*[scale=0.35]{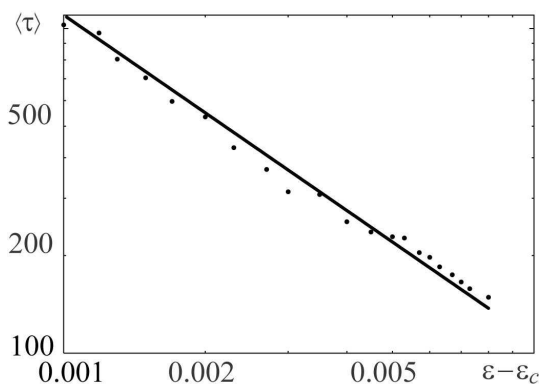}} \caption{The
log--log plot of the mean laminar phase length
$\langle\tau\rangle$ vs. the deviation
$(\varepsilon_c-\varepsilon)$ of coupling strength from the
critical value and its approximation by
$\langle\tau\rangle\sim(\varepsilon_c-\varepsilon)^{-1}$. The
power low with critical exponent $-1$ has been observed
\label{fgr:MeanTauIGS}}
\end{figure}

\begin{figure}[tb]
 \centerline{\includegraphics*[scale=0.7]{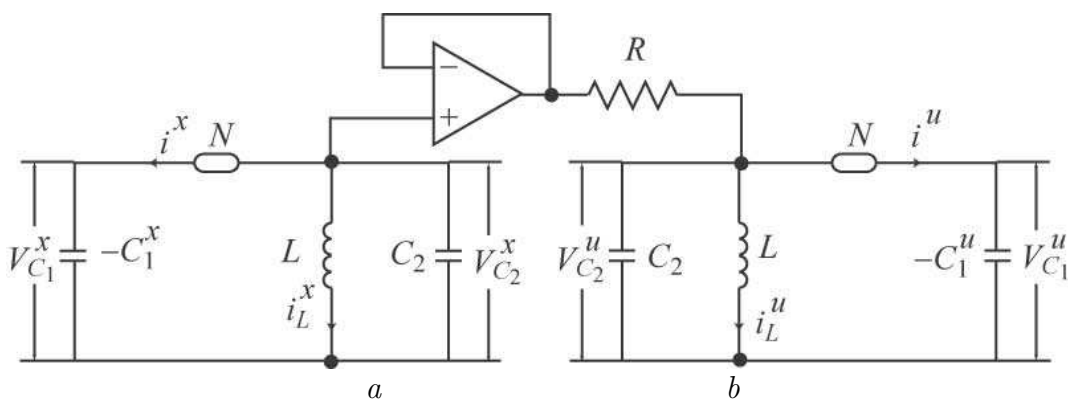}}
\centerline{\large\quad\textit{a}\qquad\qquad\qquad\qquad\qquad\quad\textit{b}}
\caption{Circuit realization of two unidirectionally coupled
Chua's oscillators \label{fgr:Scheme}}
\end{figure}

This type of behavior demonstrates the features of on--off
intermittency~\cite{Platt:1993_intermittency,Heagy:1994_intermittency}
which usually appears at the desynchronization of two coupled
identical chaotic oscillators. To analyze the statistical
properties of IGS we numerically compute the distribution
$N(\tau)$ of the laminar phases length $\tau$. We suppose that the
current state is the laminar phase if the difference between
variables $u_1(t)$ and $v_1(t)$ is less than $\Delta=0.1$,
otherwise the current state assumes to be an irregular burst.

Distribution of the laminar phases length has been shown in
Fig.~\ref{fgr:DistribIGS}. One can see that this distribution is
close to the power law with exponent ${n=-3/2}$. It should be
noticed that this result does not sensitively depend on the value
of threshold $\Delta$.

The other criteria of the on--off intermittency presence is the
dependence of the mean laminar phase length $\langle\tau\rangle$
on the deviation of coupling strength $\varepsilon$ from the
critical value $\varepsilon_c$. Fig.~\ref{fgr:MeanTauIGS} shows
the numerically determined mean laminar phase length
$\langle\tau\rangle$ versus deviation
$(\varepsilon_c-\varepsilon)$. From this figure one can see the
universal power low $\langle\tau\rangle\sim
(\varepsilon_c-\varepsilon)^k$ with critical exponent $k=-1$.
Therefore, the IGS appears to be an on--off intermittency. It is
important to note, that the ILS is also the on--off intermittency
regime (see for detail~\cite{Zhan:2002_ILS}).

It is important to note that the attractor of the considered
R\"ossler system is coherent for the selected values of the
control parameters. At the same time we have observed the IGS
presence for the different values of the control parameters when
the system behavior is characterized by the funnel attractor (see,
e.g., parameter values pointed in
\cite{Osipov:2003_3TypesTransitions}). The character of chaotic
attractor does not seem to be important for IGS as well as the
parameter mismatch influence: IGS may be observed both for small
and large mistuning of the oscillators parameters.

\begin{figure}[tb]
\centerline{\includegraphics*[scale=0.35]{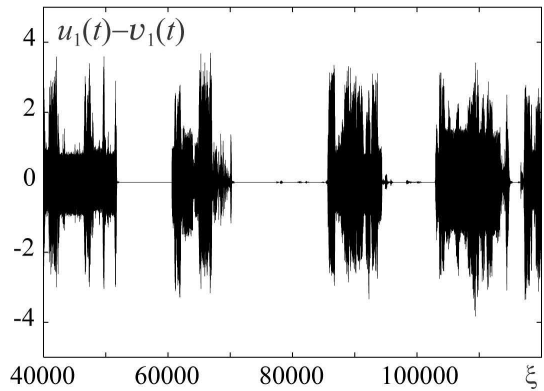}} \caption{The
dependence of difference $u_1(t)-v_1(t)$ between the coordinates
of response and auxiliary Chua's circuits on dimensionless time
$\xi$. The coupling parameter between drive and response systems
has been selected as $\varepsilon=0.03105$, the control parameters
are $\alpha_x=2.78$, $\alpha_u=2.89$, $\gamma=3.00$
\label{fgr:TorusIGS}}
\end{figure}

The intermitted generalized synchronization regime seems to appear
in the different coupled dynamical systems. To examine this
assumption we have considered the dynamics of two unidirectionally
coupled Chua's
circuits~\cite{Matsumoto:1987_TORUS,Chua:1992_Genesis} with
piecewise linear characteristic (see Fig.~\ref{fgr:Scheme}). The
typical property of the considered Chua's circuit is the presence
of two distinct characteristic time scales. Therefore, this system
may demonstrate periodical, quasi-periodical and chaotic
oscillations depending on the values of the control parameters.
The behavior of the considered Chua's circuits in the autonomous
regime is chaotic due to the selection of the control parameter
values.

The behavior of the drive Chua's circuit is described by
\begin{equation}
\begin{array}{l}
\displaystyle \dot x_{1}=-\frac{\alpha_{x}}{\gamma}f(x_{2}-x_{1}),\\
\displaystyle \dot
x_{2}=-\frac{1}{\gamma}\left(f(x_{2}-x_{1})+x_{3}\right),\\
\displaystyle\dot x_{3}=\gamma x_{2}, \label{eq:Torus_drive}
\end{array}
\end{equation}
and the equations of the response system are the following
\begin{equation}
\begin{array}{l}
\displaystyle \dot u_{1}=-\frac{\alpha_{u}}{\gamma}f(u_{2}-u_{1}),\\
\displaystyle \dot
u_{2}=-\frac{1}{\gamma}\left(f(u_{2}-u_{1})+u_{3}\right)+
\frac{\varepsilon}{\gamma}(x_{2}-u_{2}),\\
\displaystyle\dot u_{3}=\gamma u_{2}. \label{eq:Torus_responce}
\end{array}
\end{equation}

The variables $\{x,u\}_{1}=V_{C_1}^{x,u}/E$ and
$\{x,u\}_{2}=V_{C_2}^{x,u}/E$ are dimensionless voltages on
capacitors $C_1^{x,u}$ and $C_2$ of the first (drive) and the
second (response) oscillators, respectively. The variables
$\{x,u\}_{3}=i^{x,u}_L/I$ are the dimensionless currents through
the inductances $L$ in the drive and response circuits (see
Fig.~\ref{fgr:Scheme} for detail). The parameters $E$ and $I$ are
the normalization factors. Dimensionless control parameters are
$\alpha_{x,u}=C_2/C_1^{x,u}$ and
$\gamma=\frac{1}{m_1}\sqrt{{C_2}/{L}}$; $\xi=t/\sqrt{LC_2}$ is
dimensionless time. The function
\begin{equation}
f(\varsigma)=-\frac{m_0}{m_1}\varsigma+\frac{1}{2}
\left(\frac{m_0+m_1}{m_1}\right)\left(|\varsigma+1|-|\varsigma-1|\right),
\end{equation}
is the dimensionless piecewise linear voltage-current
characteristic of nonlinear element $N$, where $m_0$ and $m_1$ are
the conductivities of the corresponding branches of
voltage-current characteristic. The ratio $m_0/m_1$ has been
selected as $1/2$. The coupling parameter
${\displaystyle\varepsilon={1}/{Rm_1}}$ determines the influence
of the drive Chua's circuit on the response one.

In~Fig.~\ref{fgr:TorusIGS} the dependence of difference between
the dimensionless voltages on capacitors $C_1^{u,v}$ of response
and auxiliary system on dimensionless time $\xi$ has been shown.
One can see the presence of intermitted generalized
synchronization between the drive and response Chua's circuits.

Let us briefly discuss the nature of IGS. For this purpose we
consider the extended  phase space
$\mathbb{D}\oplus\mathbb{R}\oplus\mathbb{A}$ consisting of the
phase space of the drive ($\mathbb{D}$), response ($\mathbb{R}$)
and auxiliary $\mathbb{A}$ systems. We can also consider the
subspaces $\mathbb{D}\oplus\mathbb{R}$ and
$\mathbb{R}\oplus\mathbb{A}$ of
$\mathbb{D}\oplus\mathbb{R}\oplus\mathbb{A}$. The presence of GS
between the drive and response systems implies the complete
synchronization between the response and auxiliary systems. As the
criterion of GS one can consider the negativity of the values of
the conditional Lyapunov exponents (CLEs) \cite{Pecora:1991_ChaosSynchro,%
Pyragas:1996_WeakAndStrongSynchro} in the
$\mathbb{D}\oplus\mathbb{R}$ phase subspace while the criterion of
CS is the negativity of the transversal Lyapunov exponents (TLEs)
\cite{Anshchenko:2001_SynhroBook} in the
$\mathbb{R}\oplus\mathbb{A}$ subspace. Following the ideas of
\cite{Pyragas:1996_WeakAndStrongSynchro} one can draw a conclusion
that CLEs (in $\mathbb{D}\oplus\mathbb{R}$) and TLEs (in
$\mathbb{R}\oplus\mathbb{A}$) coincide with each other. Therefore,
when GS is destroyed in $\mathbb{D}\oplus\mathbb{R}$ with coupling
parameter decreasing, the blowout
bifurcation~\cite{Ashwinyz:1996_Bubling,Venkataramani:1996_BubblingTrans}
takes place in the $\mathbb{R}\oplus\mathbb{A}$ subspace. At the
same time it is well known that on--off intermittency takes place
below the coupling parameter value corresponding to the blowout
bifurcation. So, the on--off intermittency has to appear in the
$\mathbb{D}\oplus\mathbb{R}$ subspace below the threshold of the
GS arising.

Taking into account this idea one can draw another conclusion
concerning the noise influence. As the noise results in the
bubbling in the case of the identical chaotic oscillators for the
coupling parameter values above the point of blowout bifurcation,
the similar effects will be probably observed for GS. It means
that the intermitted behavior may be observed in the experiments
for the coupling parameter values corresponding to the GS regime
due to the noise influence.

In conclusion, we have detected the new behavior type of
unidirectionally coupled chaotic oscillators near to the onset of
generalized synchronization. This type of behavior has been called
intermitted generalized synchronization. It should be noted that
the similar effect has been observed for the unidirectionally
coupled time--delayed chaotic
oscillators~\cite{Zhan:2003_time_delayIGS}. As GS phenomenon could
also be found in the non-oscillatory chaotic systems (see, for
detail the monographic review
\cite{Boccaletti:2002_SynchroPhysReport} dealing with
synchronization phenomena in chaotic systems) IGS can also be
found in such systems below the threshold of GS arising,
correspondingly.

The intermitted generalized synchronization seems to be a common
phenomenon in nonlinear systems, and two unidirectionally coupled
chaotic oscillators usually undergo the transition to the
generalized synchronization through the on--off intermittency. The
intermitted behavior of coupled chaotic oscillators close to onset
of PS, LS and GS is likely to be the manifestation of
universalities underlying the synchronization phenomenon.

\acknowledgments We thank Svetlana V. Eremina for the English
language  support. We thank the referees for providing very
helpful comments and advices.

This work has been supported by U.S.~Civilian Research \&
Development Foundation for the Independent States of the Former
Soviet Union (CRDF, grant {REC--006}), Russian Foundation of Basic
Research (project 05--02--16273), the Supporting program of
leading Russian scientific schools (project NSch-1250.2003.2) and
the Scientific Program ``Universities of Russia'' (project
UR.01.01.371). We also thank ``Dynastiya'' Foundation.

\end{document}